# Magnetoelectric surface acoustic wave resonator with ultrahigh magnetic field sensitivity

*Liang Huang, Dandan Wen, Zhiyong Zhong, Huaiwu Zhang and Feiming Bai\**

[1]State Key Laboratory of Electronic Thin Film and Integrated Devices, University of Electronic Science and Technology, Chengdu, Sichuan, 610054, China

ABSTRACT. A magnetoelectric surface acoustic wave (MESAW) type device based on piezoelectric/magnetostrictive heterostructure was proposed to use as weak magnetic field sensor. Unlike conventional magnetoelectric bulk laminates or film stacks collecting magnetic field induced electrical charge, the MESAW detects the shift of the center frequency, which highly depends on the magnetic field due to the giant ΔE effect and the phase velocity dispersion. The magnetic field sensitivity can reach $10^{-11}$ Tesla in consideration of 100 Hz frequency accuracy. Additionally, the unique working mechanism of MESAW allows a broadband detection of weak magnetic field even with no bias magnetic field.

\*Email: fmbai@uestc.edu.cn

# I. Introduction

In the past decade, strong magnetoelectric (ME) coupling has been demonstrated in heterostructured magnetostrictive/piezoelectric laminates and films.[1–2] A number of devices based on the ME effect have been proposed: these include spintronic nonvolatile memory, gyromagnetic-field sensor, transformer, energy harvester, electrical field tunable RF/microwave devices, etc.[3-5] Detection of weak magnetic fields of pico Tesla has been achieved using magnetoelectric sensors, and in the case of the push-pull laminate, the extreme enhancement $10^{-15}$ Tesla/Hz$^{1/2}$ has been reported at the electromechanical resonance frequency (EMR).[6-7] However, giant magnetoelectric coupling and low noise density required for magnetic field sensitivity were only achievable in bulk magnetostrictive/piezoelectric laminates. Although a high ME coefficient of 737 V/cmOe was recently reported in thin film cantilever-type AlN/ FeCoSiB ME heterostructure at the EMR frequency, the low voltage from thin AlN film and the parasitic capacities cause significant magnetic noise density and low magnetic field sensitivity.[8-9] In addition to the difficulty in miniaturization of ME laminates, it turns out very challenging in sensing weak DC and low frequencies magnetic fields for both ME bulk laminates and film stacks due to the relatively low ME coefficient and the large 1/f noise.[10-11]

We here study a magnetoelectric surface acoustic wave (MESAW) type device based on similar piezoelectric/magnetostrictive heterostructure. Instead of collecting magnetic field induced electrical charge which decreases with shrinking the volume of ME sensor, the MESAW detects the shift of the center frequency with the change of magnetic field. Due to the very high working frequency of MESAW up to GHz, a broad band magnetic field can be detected. In addition, the MESAW can work with no bias magnetic field.

## II. Mechanism of MESAW

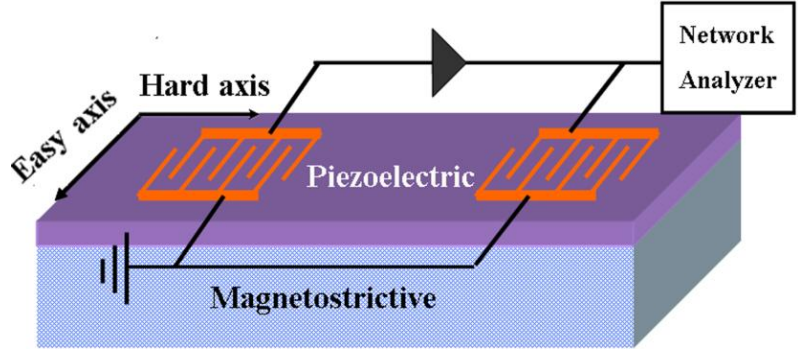

Fig. 1 Schematic illustration of MESAW structure. The SAW propagation is along the hard axis of Metglas to suppress hysteresis and ensure good linearity.

The structure of MESAW is schematically shown in Fig.1, which includes an amorphous Metglas ribbon with typical thickness of 25μm, an adhesive thin layer of Cr or Ti (not shown), a piezoelectric layer of ZnO and the very top interdigital electrodes. The Metglas substrate needs to be annealed in a transverse magnetic field, and the SAW propagation is along the hard axis of Metglas to suppress hysteresis and ensure good linearity. The center frequency of a SAW resonator is determined by

$$f_0 = v_R / \lambda \tag{1}$$

where $v_R$ is the Rayleigh wave velocity, and $\lambda$ is the wavelength of SAW. The Rayleigh wave velocity in bulk piezoelectric material is a constant, but it becomes high dispersive in layered structures. Previously, it has been demonstrated that the SAW velocity can be dramatically increased by insertion of a diamond layer with much higher Young's modulus between the piezoelectric layer and the substrate.[12-13] In current MESAW, we will show that one can change the center frequency of SAW by varying the Young's modulus ($E$) of the neighboring Metglas substrate. Upon changing magnetic field, the frequency sensitivity can be described by

$$\frac{df}{dH} = \frac{df}{dv}\frac{dv}{dE}\frac{dE}{dH} = \frac{1}{\lambda}\frac{dv}{dE}\frac{dE}{dH} \qquad (2)$$

The equation (2) gives us two clues to design and enhance the theoretic sensitivity of MESAW, i.e. increasing d$v$/d$E$ and selecting soft magnetic materials with giant ΔE effect. It is well known that Metglas ribbon is very soft and has giant ΔE effect up to 100-150%,[14] therefore, it is selected as substrate in current work. It is worth noting that the magnetostriction plays an insignificant role in the change of $f_0$, since the distance of the interdigital electrode is only slightly altered in the order of *ppm* even if a saturated magnetic field is applied. In addition, a similar multilayer structure like Fig.1 has been reported in an early work by Smole and his coauthor in an effort to make frequency tunable SAW resonator.[15] However, due to much lower Δ$E$ effect (~30%) and d$E$/d$H$ (~11GPa/Oe) of Metglas film,[16] relatively large magnetic field was needed to tune the frequency, therefore, not suitable high sensitive magnetic field sensor.

**III. Model and Computational Method**

Theoretical study of SAW characteristics with multilayered structure can be performed by the conventional approach using Campbell's method or the transfer matrix approach.[12-13] However, these methods are only applicable to piezoelectric or dielectric layers. Recently, Reinhardt and his coauthors extended the scattering matrix method to metals,[17] so the conductive Metglas layer in our case can be taken into account. In current work, we adopt this method. ZnO was treated as a polycrystalline oriented layer of finite thickness $H_{ZnO}$ on a half space Metglas with its *c*-axis normal to the interface. The elastic, piezoelectric and dielectric properties of ZnO films used for calculation come from Ref.[18].

We consider a Metglas half space covered with a ZnO layer of finite thickness $W_{ZnO}$ shown in Fig.2. The surface acoustic wave propagates along the $x_1$ axis with the $x_3$ axis normal to the layer surface. Each layer is assumed to be homogeneous. In the calculations, the aperture of the IDT along the $x_2$ axis is assumed to be infinite, then the slowness $s_2$ is set to zero. All physical quantities have a dependency of the form $\exp[j\omega(t-s_1x_1-s_3x_3)]$, where $s_1$ and $s_3$ are the slowness along the $x_1$ and $x_3$ axis, respectively, and the angular frequency is given by $\omega=k_i/s_i (i=1,3)$.

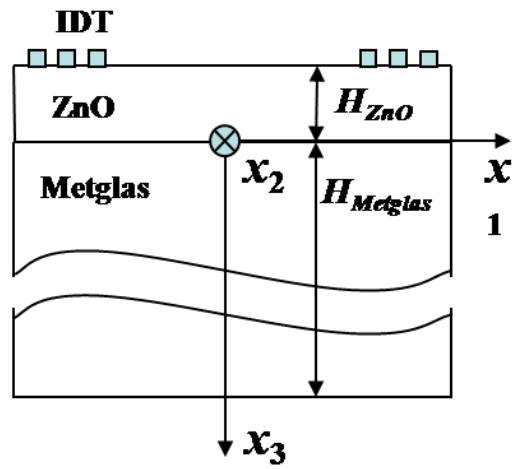

Fig.2. Schematic of the ZnO/Metglas multilayered structure

A. Plane wave propagation

(i) Case of the ZnO layer

As we all know, the ZnO layer is an insulating piezoelectric layer, inside which both Newton's and Maxwell's equations should be satisfied. In the quasi-static approximation, they can be expressed as:

$$\frac{\partial T_{ij}}{\partial x_i} = \rho \frac{\partial^2 u_j}{\partial t^2}, \tag{3}$$

$$\frac{\partial D_i}{\partial x_i} = 0, \quad (i, j = 1, 2, 3) \tag{4}$$

5Magnetoelectric surface acoustic wave resonator with ultrahigh magnetic field sensitivity
L. Huang, D. D. Wen et al.

where $T_{ij}$ is the stress tensor, $u_j$ is the mechanical displacement, $D_i$ is the electric displacement, and $\rho$ is the mass density, respectively. Additionally, the stress tensor $T_{ij}$ and the electric displacement $D_i$ can be obtained by the equations of piezoelectricity, which can be written as

$$T_{ij} = c_{ijkl}\frac{\partial u_k}{\partial x_l} + e_{lij}\frac{\partial \phi}{\partial x_l}, \tag{5}$$

$$D_i = e_{ikl}\frac{\partial u_k}{\partial x_l} - \varepsilon_{il}\frac{\partial \phi}{\partial x_l}, \quad (i,j,k,l=1,2,3) \tag{6}$$

where $\phi$ is the electric potential, $c_{ijkl}$ is the elastic tensor, $e_{lij}$ is the piezoelectric tensor, and $\varepsilon_{ij}$ is the dielectric tensor, respectively.

From Eq. (3) to (6), we get

$$M\mathbf{h} = s_3\mathbf{h}, \tag{7}$$

in which $\mathbf{h} = (u_1, u_2, u_3, \phi, -\frac{T_{13}}{j\omega}, -\frac{T_{23}}{j\omega}, -\frac{T_{33}}{j\omega}, -\frac{D_3}{j\omega})^T$ is the eight-component state vector that can fully describe the distribution of the electromechanical fields in the ZnO layer, and the 8×8 matrix $M$ is given by

$$M = \begin{bmatrix} -s_1 A_{33}^{-1} A_{31} & A_{33}^{-1} \\ -s_1^2(A_{11} - A_{13}A_{33}^{-1}A_{31}) + P & -s_1 A_{13}A_{33}^{-1} \end{bmatrix}, \tag{8}$$

where the $A_{il}$ matrices are composed of the material constants:

$$A_{il} = \begin{bmatrix} c_{i1l1} & c_{i12l} & c_{i13l} & e_{li1} \\ c_{i21l} & c_{i22l} & c_{i23l} & e_{li2} \\ c_{i31l} & c_{i32l} & c_{i33l} & e_{li3} \\ e_{i1l} & e_{i2l} & e_{i3l} & -\varepsilon_{il} \end{bmatrix}, \tag{9}$$

and the matrix P has the form

$$P = \begin{bmatrix} \rho & 0 & 0 & 0 \\ 0 & \rho & 0 & 0 \\ 0 & 0 & \rho & 0 \\ 0 & 0 & 0 & 0 \end{bmatrix}, \tag{10}$$

respectively.

Once the eigenvalues and eigenvectors of the matrix $M$ have been calculated, solutions of Eq. (7) can be obtained as

$$\mathbf{h} = F\Delta(x_3)\mathbf{a}\exp[2j\pi f(t - s_1 x_1)] \tag{11}$$

where $F$ is a 8×8 matrix composed of the eight vertically arranged eigenvectors, $\Delta(x_3) = diag[\exp(-2j\pi f s_3^{(m)} x_3)]$, m = 1, 2, …, 8, is a 8×8 diagonal matrix describing the dependence along the $x_3$ axis through the eight corresponding eigenvalues $s_3^{(m)}$, and $\mathbf{a}$ is a vector of the amplitudes of the partial waves that can be obtained when the boundary conditions are specified.

(ii) Case of the Metglas semi-infinite substrate

The Metglas is a metallic layer, inside which acoustic propagation is not coupled to electromagnetic fields, so Eq. (5-6) is no longer valid. The only two equations that can satisfy this case are Hooke's law:

$$T_{ij} = c_{ijkl}\frac{\partial u_k}{\partial x_l}, \tag{12}$$

and Newton's law in Eq. (1). Consequently, the state vector should be rewritten as

$$\mathbf{h}' = (u_1, u_2, u_3, -\frac{T_{13}}{j\omega}, -\frac{T_{23}}{j\omega}, -\frac{T_{33}}{j\omega})^T. \tag{13}$$

The matrices $A'_{il}$ currently only include the elastic constants of the Metglas layer with the form

$$A'_{il} = \begin{bmatrix} c_{i11l} & c_{i12l} & c_{i13l} \\ c_{i21l} & c_{i22l} & c_{i23l} \\ c_{i31l} & c_{i32l} & c_{i33l} \end{bmatrix}, \tag{14}$$

and the matrix including the mass density becomes

$$P' = \begin{bmatrix} \rho & 0 & 0 \\ 0 & \rho & 0 \\ 0 & 0 & \rho \end{bmatrix}, \qquad (15)$$

respectively. For this reason, only six eigenvalues and six corresponding eigenvectors can be obtained with the dimension of the matrix $M$ shrinking from eight to six.

B. Classification of the partial waves

The eigenvalues of the matrix $M$ represented partial waves are either pair of conjugate complex or opposite real values. It is extremely important to sort the partial waves into two groups according to the classification rules summarized by Table 1.[17]

Table 1. Partial modes classification rules.

| Partial modes | Inhomogeneous waves | Propagating waves |
| --- | --- | --- |
| Reflected(+) | $\Im(s_3^{(m)}) < 0$ | $P_3^{(m)} > 0$ and $\Im(s_3^{(m)}) = 0$ |
| Incident(-) | $\Im(s_3^{(m)}) > 0$ | $P_3^{(m)} < 0$ and $\Im(s_3^{(m)}) = 0$ |

* $\Im(s_3^{(m)})$ is the imaginary part of the eigenvalues, and $P_3^{(m)}$ is the vertical component of the Poynting vector with the partial mode number $m$.

*For the ZnO layer, $m = 1, 2, 3,\ldots,8$; For the semi-infinite Metglas substrate, $m = 1, 2, 3,\ldots,6$.

C. Scattering matrix algorithm

In order to discuss conveniently, an auxiliary variable is introduced in the following for each layer, which is defined by

$$g(x_3) = \begin{bmatrix} g^{(+)}(x_3) \\ g^{(-)}(x_3) \end{bmatrix}$$
$$= \begin{bmatrix} \Delta^{(+)}(x_3) & 0 \\ 0 & \Delta^{(-)}(x_3) \end{bmatrix} \begin{bmatrix} \mathbf{a}^{(+)} \\ \mathbf{a}^{(-)} \end{bmatrix}, \qquad (16)$$

where the reflected(+) partial waves are stored in the matrix $\Delta^{(+)}(x_3)$ and the vector $\mathbf{a}^{(+)}$, while the incident(-) partial waves are stored in the $\Delta^{(-)}(x_3)$ and $\mathbf{a}^{(-)}$. Then, at any position $x_3$ inside one layer, the $g(x_3)$ can be expressed only as a function of $g^{(+)}(x_3)$ as follows:

$$g(x_3) = \begin{bmatrix} I_d \\ R(x_3) \end{bmatrix} g^{(+)}(x_3), \tag{17}$$

where $d = 3$ or $4$ for the Metglas layer or the ZnO layer, and $R(x_3)$ with dimension of rank $3\times3$ or $4\times4$ is a reflection matrix defined by

$$g^{(-)}(x_3) = R(x_3)g^{(+)}(x_3). \tag{18}$$

Denoting $x_3$ and $x_3^{'}$ two positions in one layer, from Eq. (15) and (17) we can get

$$\begin{aligned} g^{(-)}(x_3^{'}) &= \Delta^{(-)}(x_3^{'} - x_3)g^{(-)}(x_3) \\ &= \Delta^{(-)}(x_3^{'} - x_3)R(x_3)\Delta^{(+)}(x_3 - x_3^{'})g^{(+)}(x_3^{'}), \end{aligned} \tag{19}$$

so, $R(x_3^{'}) = \Delta^{(-)}(x_3^{'} - x_3)R(x_3)\Delta^{(+)}(x_3 - x_3^{'})$. \hfill (20)

If the positions $x_3$ and $x_3^{'}$ are at the bottom and top of one layer respectively, we can calculate the reflection matrix at the top of one layer according to

$$R^t = \Delta^{(-)}(-W)R^b\Delta^{(+)}(W), \tag{21}$$

where $W$ is the thickness of one layer.

In our model, the Metglas substrate is a semi-infinite metal layer, no reflection occurs so that the reflection matrix of rank $3\times3$ at the bottom of the Metglas layer is $R_{Metglas}^b = 0$. Therefore, the reflection matrix at the top of the Metglas substrate can be calculated as

$$R_{Metglas}^t = \Delta^{(-)}(-W_{Metglas})R_{Metglas}^b\Delta^{(+)}(W_{Metglas}) = 0$$

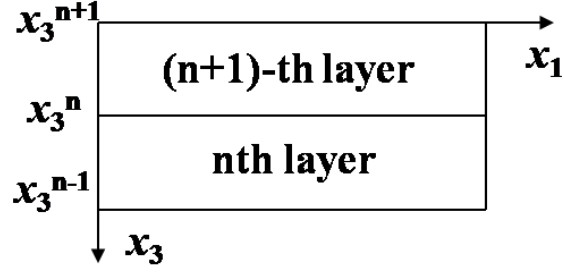

Fig.3. Schematic of purely piezoelectric or dielectric multilayered structure

In order to calculate the $R_{ZnO}{}^b$, the mode conversion is needed to be considered between the ZnO and Metglas layer. For case of purely piezoelectric or dielectric multilayered structures shown in Fig. 4, the continuity of the state vector can be directly written as

$$g_{n+1}(x_3{}^n) = F_{n+1}{}^{-1} F_n g_n(x_3{}^n) = F_{n+1}{}^{-1} F_n \begin{bmatrix} I_4 \\ R_n{}^t \end{bmatrix} g_n{}^{(+)}(x_3{}^n) = \begin{bmatrix} A \\ B \end{bmatrix} g_n{}^{(+)}(x_3{}^n) \quad (22)$$

where $x_3{}^n$ is the coordinate value in $x_3$ axis between the nth and (n+1)-th layer. From Eq. (22), we can obtain the reflection matrix at the bottom of the (n+1)-th layer as

$$R_{n+1}{}^b = BA^{-1}. \quad (23)$$

However, because the dimension of $R_{ZnO}{}^b$ is larger than that of $R_{Metglas}{}^b$, the continuity of the state vector cannot be written as the form in Eq. (22). To equilibrate the system, the electric surface charge density can be assumed to accumulate at the interface between the ZnO and Metglas layer. Then, the reflection matrix at the bottom of the ZnO layer can be calculated as

$$R_{ZnO}{}^b = FE^{-1}, \quad (24)$$

with the modification given by

$$\begin{bmatrix} E \\ F \end{bmatrix} = F_{ZnO}{}^{-1} \begin{bmatrix} C & 0 \\ 0 & 0 \\ D & 0 \\ 0 & 1 \end{bmatrix}, \text{ where } \begin{bmatrix} C \\ D \end{bmatrix} = F_{Metglas} \begin{bmatrix} I_3 \\ R_{Metglas}{}^t \end{bmatrix}. \quad (25)$$

Once again, Eq. (21) can be used to calculate the reflection matrix at the top of the ZnO layer.

D. Calculation of the effective surface permittivity

Based on the computational method discussed above, the state vector on the top surface of the ZnO/Metglas multilayered structure can be given by

$$\mathbf{h}(-w_{ZnO}) = F_{ZnO} \begin{bmatrix} I_4 \\ R_{ZnO}^t \end{bmatrix} g_{ZnO}^{(+)}(-w_{ZnO}) = \begin{bmatrix} N \\ P \end{bmatrix} g_{ZnO}^{(+)}(-w_{ZnO}) \qquad (26)$$

in which the two sub-matrices $N$ and $P$ are associated with the generalized displacement and stress, respectively. In consideration of the electric behavior of the vacuum, the eigenvector matrix $F_{ZnO}$ in Eq.(26) must be modified according to the rule

$$F_{8,i} \leftarrow F_{8,i} + j\varepsilon_0 |s_1| F_{4,i}, \quad i = 1,2,3,\cdots,8. \qquad (27)$$

From Eq. (26), we can easily get the surface Green's function according to

$$G = NP^{-1} \qquad (28)$$

Additionally, it is possible to obtain the surface effective permittivity through

$$\varepsilon_{eff}(f, s_1) = \frac{1}{j|s_1|G_{44}(f, s_1)} \qquad (29)$$

**IV. Results and discussion**

Fig.4 shows the effective permittivity of a ZnO/Metglas half space as a function of the phase velocity and the thickness of ZnO layer. In the calculation, the frequency was chosen as 0.7 GHz and 1.3 GHz, Fig.4(a) and Fig.4(b), respectively. As shown in Fig.4(a), the pole indicates the surface wave solution for a metalized surface, because it gives a finite charge density at zero electric potential. The zero corresponds to the surface wave solution for a free surface, because the charge density is zero. In the corresponding 2D views shown in Fig.4(c), the white line represents the calculated $H_{ZnO}$ dependence of phase velocities. Only the basic Rayleigh mode was found. This is due to the stiffening of the substrate by the ZnO layer. The velocity increases smoothly with increasing $H_{ZnO}$ from 0.1 μm to 1.2 μm when the frequency is set as 0.7 GHz, as shown in Fig. 4(c). But there is a broken line between 0.7 and 2.7 μm on Fig. 4(d) when the frequency was set as 1.3 GHz, indicating that the Rayleigh wave cannot be excited. No curvature bend was found for $H_{ZnO}$ between 2.7 μm and 4 μm, as shown in Fig. 4(d).

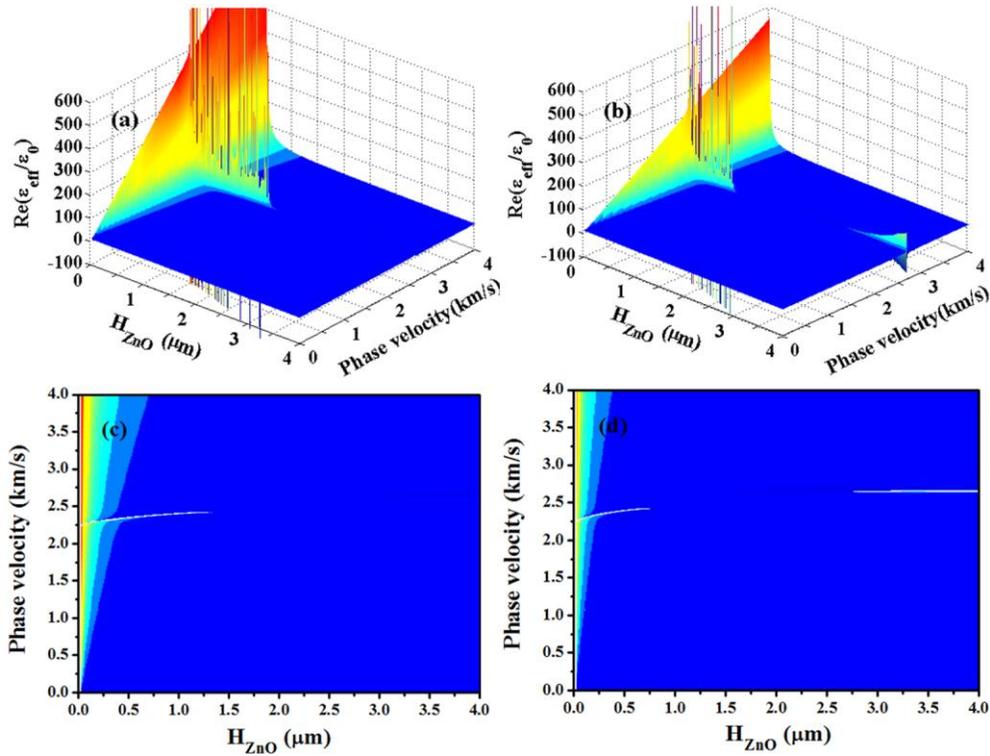

Fig.4 The effective permittivity of a ZnO/Metglas half space as a function of the phase velocity and the thickness of ZnO layer with (a) the center frequency of 0.7GHz and (b) the center frequency of 1.3GHz. (c) and (d) the corresponding 2D views of calculated $H_{ZnO}$ dependence of phase velocities.

The electro-mechanical coupling coefficient $K^2$ values can be calculated from the equation $K^2 = 2(v_0 - v_s)/v_0$, where $v_0$ and $v_s$ are the velocities with the boundary where IDT is set electrically open and shorted, respectively. Fig.5 shows the calculated $K^2$ dependence on $H_{ZnO}$ of the ZnO/Metglas half space. Here the frequency was selected at 1.3 GHz. It can be seen that the $K^2$ value first increases then decreases. For the thickness of 0.05 to 0.5 µm, the $K^2$ value is above 0.5%, fourfold that of 90° cut quartz, and relatively stable. We also calculated the $K^2$ value for a ZnO layer thicker than 3 µm. As shown in Fig. 5(b), the $K^2$ value increases with increasing the thickness of ZnO layer, and almost reach 1%. But it was found the SAW velocity of thicker ZnO layer is less dependent on the change of Young's modulus of Metglas half space. So in the following discussion, thin ZnO layer of 0.3 µm thickness was used.

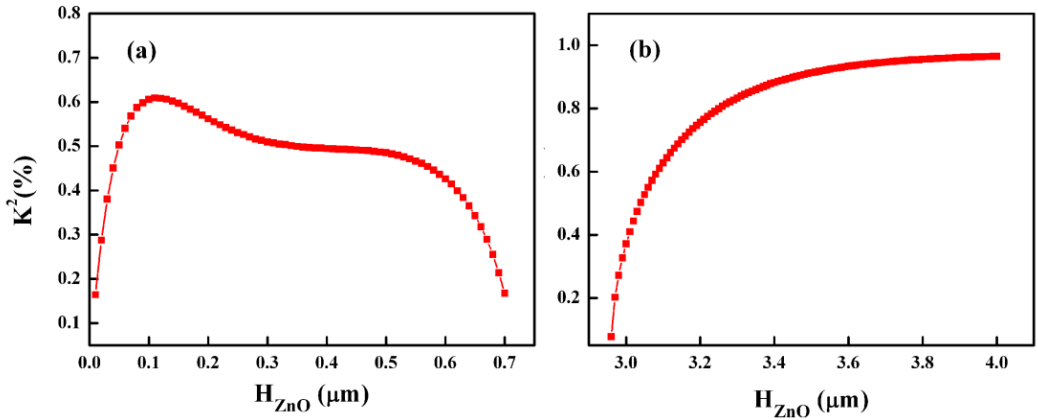

Fig. 5 Calculated $H_{ZnO}$ dependence of $K^2$ of the ZnO/Metglas half space, (a) low ZnO thickness and (b) high ZnO thickness.

Fig.6 shows the frequency response of IDT transducer on the ZnO/Metglas half space based on the effective permittivity model, which is more powerful in revealing the reduction of null frequency bandwidth and the insertion loss than the conventional delta function or the dispersive delta function models.[14]. The center frequency was chosen at 1.3 GHz and the numbers of electrode pairs were 40 and 100. The calculated insertion loss are -14 dB and -6 dB for 40 electrode pairs and 100 pairs, respectively. Obviously, large pair number is not only helpful to reduce the null frequency bandwidth and thus enhance magnetic field sensitivity, but also to reduce the insertion loss to -6dB, which is much smaller than the reported value of quartz, and sufficiently low for general SAW resonator.

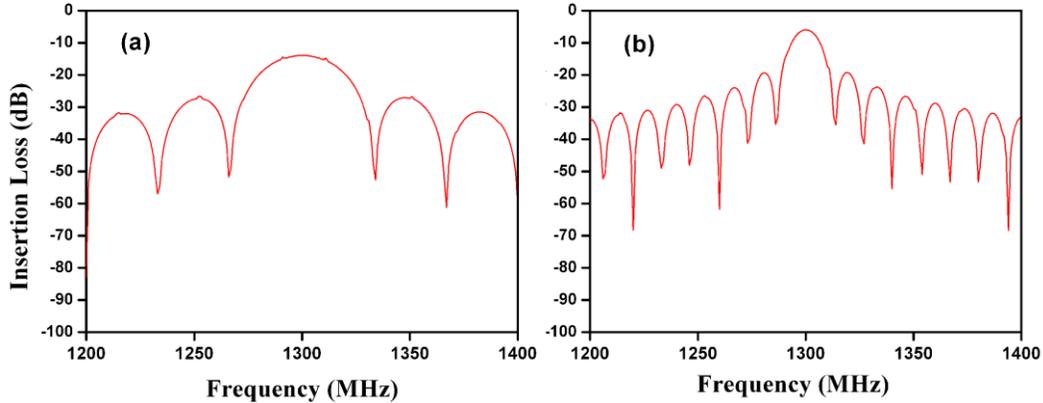

Fig. 6 Frequency response of IDT transducer on the ZnO/Metglas half space with different number of electrodes, (a) N=40 and (b) N=100. The center frequency was chosen at 1.3GHz.

Fig. 7 shows the dependence of SAW velocity on the change of Young's modulus of Metglas half space. Clearly, with increasing E from 80 to 140 GPa, the phase velocity increases from ~2050 to 2600 m/s, close to the bulk phase velocity of ZnO. On the contrary, the d$v$/d$E$ decreases from $13 \times 10^{-9}$ to $8 \times 10^{-9}$ m$^3$/(N.s). We have also calculated the change of center frequency of SAW, which significantly shifted from 1.12 GHz to 1.44 GHz upon varying E from 80 GPa to 140 GPa (not shown). Such dramatic change of Young's modulus is solid based on previous studies of the magnetomechanical coupling and ΔE effect of Metglas by Anderson.[14] Here, the inset of Fig.7 are replotted from the experimental results in reference [14], where Metglas 2605S-3 ($Fe_{19}Si_5B_{16}$) was annealed for 30 minutes in a transverse field. For the convenience, the d$E$/d$H$ is also shown in the same figure, which determines the frequency sensitivity of MESAW. As can be seen, the d$E$/d$H$ is not zero at H=0, indicating that the MESAW can work without magnetic bias. In addition, two maximal d$E$/d$H$ values ~150 GPa/Oe are obtained under bias magnetic fields of 0.5 Oe and 1.1 Oe, corresponding to the highest magnetic field sensitivity. Fig.7 allows us to estimate the optimal frequency sensitivity upon applying magnetic field. Taking 1.1 Oe magnetic biasing field for example, very high frequency shift about 805 MHz/Oe can be achieved. Of course, the measurement of frequency is critically affected by temperature compensation and frequency counter, which can be improved by analogue and digital designs. In practical, 10 Hz frequency accuracy was obtained in SAW-based force or pressure sensors working in the frequency range of 200~400 MHz.[19] In consideration of 100 Hz frequency accuracy of current 1.3 GHz MESAW sensor, ultrahigh magnetic field sensitivity of ~$10^{-11}$ Tesla can be obtained.

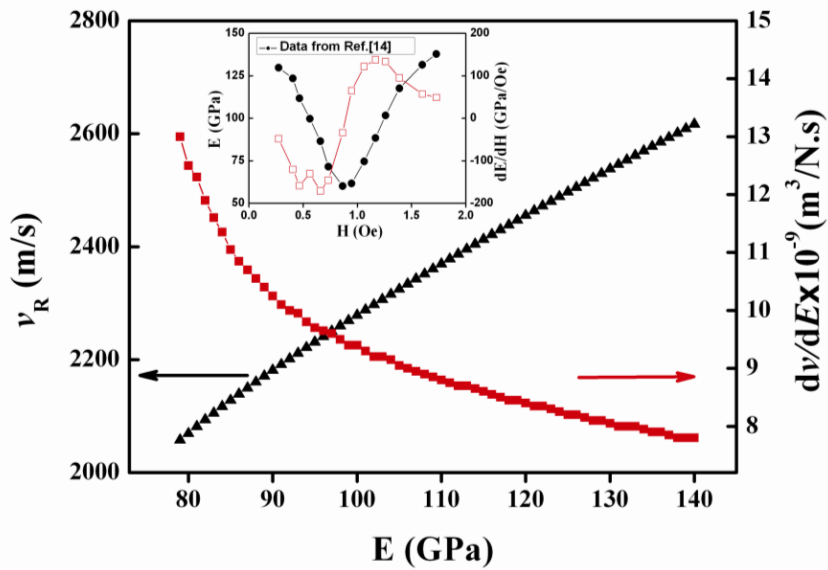

Fig. 7 Dependence of SAW velocity and d$v$/d$E$ on the change of Young's modulus of Metglas half space. The inset shows the Young's modulus and $d$E/$d$H versus applied bias field for Metglas 2605S3 annealed for 30 minutes in a transverse field. Data replotted from reference [14].

It is realized that the sensitivity of the MESAW crucially depends on whether or not high quality ZnO film can be grown on the Metglas ribbon without degradation of magnetic properties. We have directly grown 600nm-thickness ZnO film on the commercial Metglas 2605SC ribbon ($Fe_{81}Si_{3.5}B_{13.5}C_2$) at 300°C in Ar by RF magnetron sputtering. Fig. 8(a) showed that only ZnO (002) peak can be seen from the X-ray diffraction analysis, and the full width at half maximum (FWHM) of the $(002)_{ZnO}$ peak is only 4.28°, indicating excellent texture. No crystalline iron peak was found. Fig. 8(b) shows the magnetization versus the applied magnetic field (M-H) curves. Almost no difference of magnetic properties was found between the as-received Metglas ribbon and the one after deposition of the ZnO film, meaning that Metglas ribbon was well preserved.

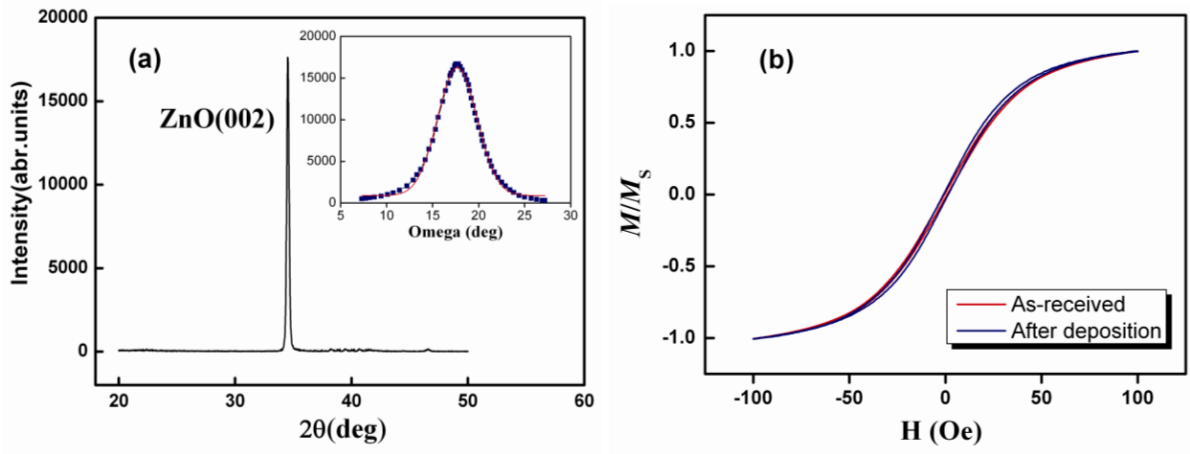

Fig.8 (a) XRD line scan of ZnO film deposited on Metglas 2605SC, (b) the M-H curves of the as-received Metglas ribbon and the one after deposition of the ZnO film. The insert of (a) shows the rocking curve of ZnO (002) peak.

## V. Conclusion

In summary, we suggest that a magnetoelectric surface acoustic wave resonator can serve as a highly sensitive weak magnetic field sensor. The dependence of phase velocity of ZnO/Metglas half space on the change of Young's modulus of metglas was calculated according to a stable scattering matrix method. It is shown that the center frequency of MESAW sensor is highly dependent on the magnetic field due to the giant ΔE effect of metglas and phase velocity dispersion. In consideration of 100 Hz frequency accuracy, ultrahigh magnetic field sensitivity of ~$10^{-11}$ Tesla can be obtained. To demonstrate the feasibility of the proposed MESAW, highly textured ZnO film was directly deposited on a commercial Metglas ribbon with well preserved magnetic properties. Since the $\Delta E$ effect is insensitive to the change of magnetic field frequency and free from ferroelectric loss,[20] the MESAW can be applied in very wide frequency, for example DC~$10^4$ Hz, therefore very useful in biomedical applications or positioning devices.

## ACKNOWLEDGMENTS

The authors would like to acknowledge financial support from the National Basic Research Program of China under Grant No. 2012CB933104 and the National Science Fund of China under grant No. 61271031.